\documentclass[runningheads]{llncs}
\usepackage[T1]{fontenc}
\usepackage{amsmath}
\usepackage{graphicx}

\usepackage{hyperref}

\begin{document}
\title{MPTopic: Improving topic modeling via Masked Permuted pre-training}
\titlerunning{MPTopic: Masked Permuted pre-training}
%
\author{Xinche Zhang\inst{1}\orcidID{0000-0002-1990-9041} \and
        Evangelos Milios\inst{1}\orcidID{0000-0001-5549-4675}}
\authorrunning{Xinche, Milios}
%
\institute{Dalhousie University, Halifax NS 08544, Canada \\
\email{xn233362@dal.ca, eem@cs.dal.ca}\\
}
\maketitle              
\begin{abstract}




Topic modeling is pivotal in discerning hidden semantic structures within texts, thereby generating meaningful descriptive keywords. While innovative techniques like BERTopic and Top2Vec have recently emerged in the forefront, they manifest certain limitations. Our analysis indicates that these methods might not prioritize the refinement of their clustering mechanism, potentially compromising the quality of derived topic clusters. To illustrate, Top2Vec designates the centroids of clustering results to represent topics, whereas BERTopic harnesses C-TF-IDF for its topic extraction.In response to these challenges, we introduce "TF-RDF" (Term Frequency - Relative Document Frequency), a distinctive approach to assess the relevance of terms within a document. Building on the strengths of TF-RDF, we present MPTopic, a clustering algorithm intrinsically driven by the insights of TF-RDF. Through comprehensive evaluation, it is evident that the topic keywords identified with the synergy of MPTopic and TF-RDF outperform those extracted by both BERTopic and Top2Vec.

\end{abstract}

\section{Introduction}

\begin{figure}
\centering
\includegraphics[width=1\textwidth]{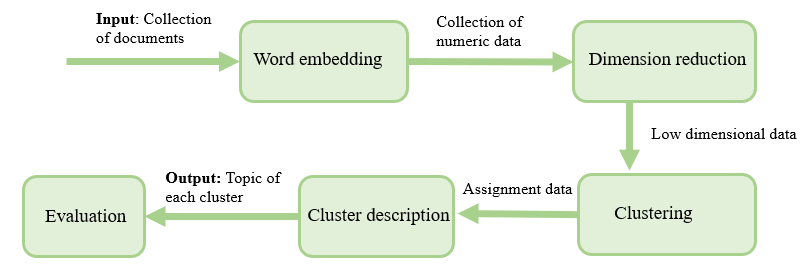}
\caption{\label{fig:flowchart}The input is the collection of documents and the output is the topic}
\end{figure}

\begin{figure}
\centering
\includegraphics[width=1\textwidth]{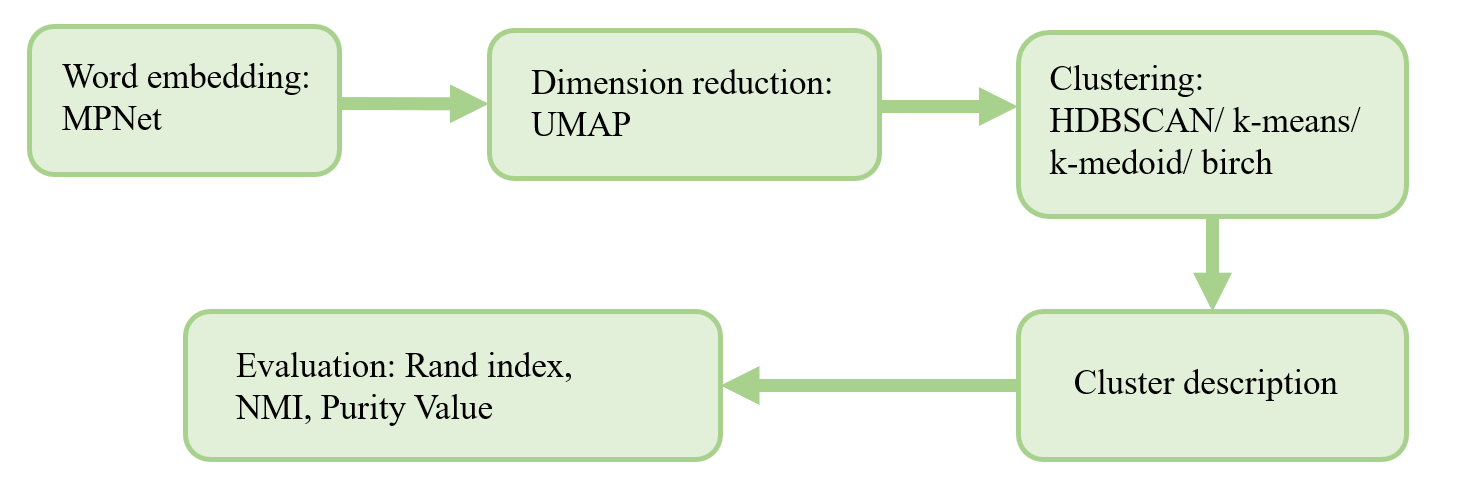}
\caption{\label{fig:flowchart2}Different techniques are used for these steps}
\end{figure}

Due to the rapid development of information technology, the number of documents people need to handle at work is becoming increasingly large, and people are gradually unable to cope with and manage a considerable amount of documents. Manually identifying thousands of documents, labeling them, and classifying them by topic seems impossible for individuals. In this case, topic modeling can traverse all texts and intelligently select topics for each document. It saves much time for the viewer. Document clustering is also an essential feature in the field of NLP, which improves efficiency by generating word embedding to classify a large number of documents according to user needs.

Nowadays, many mature and high-quality topic modeling algorithms are available on the web, such as BERTopic~\cite{bertopic}  and Top2Vec~\cite{top2vec}. They use document clustering to help the topic modeling process. However, they still have some flaws. First, these techniques have some limitations in the clustering algorithm. For example, both BERTopic and top2vec only provide HDBSCAN clustering~\cite{sklearn}. Although HDBSCAN~\cite{hdbscan} is a quality clustering algorithm, users need more options in the clustering algorithm because HDBSCAN does not work well when the user wants to specify the number of clusters. Second, the approach employed by Top2Vec, which utilizes the centroids of clusters to delineate topics, can inadvertently overlook words that, despite being relevant to a particular topic, may not significantly influence the position of the centroid. Bertopic employs the C-TF-IDF method to discern topics. However, certain mathematical inconsistencies within its formulation could render it ineffective at filtering out stop words in specific situations.

This paper proposes a new NLP algorithm named MPTopic for generating text clusters and detecting text topics based on word embeddings of MPNet~\cite{mpnet}. MPTopic will use a collection of documents from users as input, divide these data into different clusters, and generate a topic for each cluster(This is similar to Bertopic and Top2vec).   We have taken advantage of existing algorithms and improved the abovementioned flaws. Our algorithm provides multiple clustering algorithm options, such as k-means, birch~\cite{birch}, and HDBSCAN, so that users can make the most appropriate choice based on their needs. We have proved that when an estimate of cluster numbers was provided(such as 20 for the 20newsgroup dataset), using k-means instead of HDBSCAN led to a much better clustering result(purity value). In terms of cluster description, both TF-IDF and C-TF-IDF exhibit inherent limitations. While they are adept at handling smaller documents, their efficacy diminishes for larger ones. Specifically, the IDF component will be biased, making TF-IDF no loger effective at capturing key information in the document.

Compared to the existing algorithm (Top2Vec and BERTopic), the MPTopic made two main contributions: it increases the quality of clustering results so that the topic modeling result will be potentially improved. Furthermore, we implement the novel TF-RDF technique to recognize the topics in documents.

\section{Background}

In the era of big data, textual data has proliferated at an astonishing rate. From social media platforms, business reports, to scholarly articles, the overwhelming volume of textual content demands efficient and effective processing techniques to extract meaningful insights. Traditional methods of managing and analyzing textual data relied heavily on manual curation, which has become increasingly infeasible given the current data deluge. Consequently, a need for sophisticated automatic techniques that can parse, analyze, and categorize vast amounts of textual data is evident.

Topic modeling has emerged as a potent tool in addressing this challenge. By distilling vast textual corpora into coherent themes or 'topics', it provides a bird's-eye view of the primary subjects covered in the data, thereby facilitating efficient data management and knowledge discovery. The primary input for topic modeling is a collection of documents. A 'document' here can refer to anything from a short article or paragraph to longer pieces like books or research papers. The result of topic modeling is typically a set of topics. Each topic is a set of words that could together describe the general idea of a document, thereby suggesting a specific theme or subject.

Historically, methods like Latent Dirichlet Allocation (LDA) have been widely adopted for topic modeling, but with the advent of deep learning and advancements in natural language processing, newer algorithms have surfaced. Techniques like BERTopic and Top2Vec, which leverage state-of-the-art embeddings(Doc2Vec ~\cite{doc2vec} and MiniLM~\cite{minilm}) and clustering mechanisms(HDBSCAN~\cite{hdbscan}), have demonstrated enhanced performance in many scenarios. However, like all models, they are not without their drawbacks.

Top2Vec, BERTopic, and our proposed methodology all encompass a segment termed "cluster description." Within this phase, all documents with the similar content have been in one cluster. The objective now shifts to discerning an apt topic representation for each cluster, a process that is inherently dynamic. While TF-IDF and C-TF-IDF, as employed by BERTopic, offer viable solutions, their effectiveness wanes significantly with an upsurge in document size. To illustrate, considering a document of the magnitude of an entire book, or in the frameworks of BERTopic and MPTopic where entire clusters might be viewed as singular documents, the resultant size might surpass 100,000 words. Implementing TF-IDF or C-TF-IDF on such extensive documents undermines their efficacy dramatically. Consequentially, there's an inadvertent elevation of stop words (e.g., "the", "of") to the status of topics, which detracts from the quality of topic representation.

\section{Methodology}

In the methodology section, we will discuss the main components of MPTopic: MPNet word embedding, dimension reduction, clustering algorithms, topic modeling by TF-RDF and evaluation.

\subsection{MPNet Word embeddings}

In the first step of the text clustering algorithm, we first convert the text into word vectors. For a text clustering task, before using a clustering algorithm, we expect and assume that the vectors corresponding to similar texts in a series of texts are very similar in space. Thus, a good quality word embedding is often a good start for an NLP task. MPTopic uses sentence-Transformer to introduce the MPNet model~\cite{mpnet}. Sentance-transformer~\cite{st} is a framework based on Transformer and PyTorch that makes it easy to fine-tune and use pre-trained models for various tasks.


We will convert the documents input by the user into word vectors by using MPNet's word embedding pre-trained model. Each document will be represented by a 768-dimensional vector when the conversion is complete. When all documents are converted, we obtain a set of high-dimensional vectors equal to the number of documents, and we will implement a clustering algorithm on these high-dimensional vectors in the next step. 

Input from the user needs to be in the form of a list containing multiple strings. Each string is the context of a document. MPTopic receives the list and converts it to the 'numpy.ndarray' format, which contains multiple lists, each containing 768 floats representing 768 dimensions of data.

\subsection{Dimension reduction}

If the user uses the HDBSCAN algorithm, we must reduce the dimension before the clustering algorithm. MPTopic provides a variety of dimension reduction algorithms, including PCA, T-SNE, and UMP. UMAP~\cite{umap} is the default algorithm, which preserves more features of high-dimensional space in low-dimensional space as much as possible. It could show excellent performance when combined with HDBSCAN.

The main parameters of UMAP are n\_Neighbors, n\_Components, and metric. Here, our default value for them is 15, 5, and 'cosine,' respectively.

\subsection{Clustering algorithm}

This project provides various clustering algorithms, including Birch, k-medoid, k-means, and HDBSCAN. After our experiments, we recommend two of them here: k-means and HDBSCAN.
Hierarchical Density-Based Spatial Clustering of Applications with Noise (HDBSCAN) is one of NLP's most commonly used Clustering algorithms. It extends the DBSCAN algorithm by converting it into a hierarchical clustering algorithm, and it is widely used for all kinds of unsupervised text categorization tasks because it has many advantages. First, an important feature is that the HDBSCAN algorithm does not need to specify the number of targets for clustering. His main parameter is min\_cluster\_size, which means we only need to specify the minimum number of documents in a cluster, and the algorithm will automatically divide the cluster based on the spatial distribution of the vectors. Another feature of the HDBSCAN algorithm is that not all values are recognized as one of the clusters,d, and some documents are recognized as outliers. This feature is very effective for the clustering algorithm because when we cannot guarantee the quality and quantity of the input documents, there are always some documents that do not belong to any existing cluster and cannot form a new cluster. It is unreasonable to forcibly assign these documents to a particular cluster.

At the same time, we have to analyze some shortcomings of the HDBSCAN method. We think the problem is with the parameters. Two main parameters in HDBSCAN are min\_cluster\_size and min\_simple. We think these two parameters are not intuitive for users. Firstly, for min\_SAMPLE, the HDBSCAN algorithm is not sensitive to min\_simple, and the official document explains this parameter as follows: "The very simple intuition for what min\_samples does provide a measure of how conservative you want your clustering to be." We can see how this parameter is used, but it does not give us a clear idea of what value is appropriate for our data set. Another parameter is min\_cluster\_size, which is usually the only parameter we need to specify (min\_samples defaults to be equal to min\_cluster\_size), which is convenient, but it can be difficult for users to choose the correct value.
For example, for a 20Newsgroup training dataset with about 12,000 documents, theoretically, each cluster should have at least 100 documents. However, when we tried to set the min\_cluster\_size to 100, we only ended up with seven datasets. This means that every three topics here are merged into one, which is not what we want. When this happens, the user has to continuously fine-tune the parameters until the algorithm outputs a suitable result, which significantly reduces the algorithm's efficiency. Another disadvantage of HDBSCAN is that it cannot deal with data with too high dimensions. However, the output of the mainstream word embedding pre-trained technology is always hundreds of dimensions. To keep the word vector high performance in HDBSCAN, we usually have to use dimension reduction technology to reduce the hundreds of dimensions to 5-10. Although UMAP dimension reduction has been proven to fit well with HDBSCAN, dimension reduction will eventually lead to some information loss.~\cite{sklearn}

Therefore, unlike other text clustering algorithms, the MPTopic algorithm provides a variety of clustering algorithms. K-means is one of these algorithms. This fundamental algorithm is simple, but it forms a good complement to the HDBSCAN algorithm in MPTopic. First of all, different from HDBSCAN, the parameter of K-means is the number of target clusters, which applies to the situation where users have grasped the approximate K-value, i.e., they already know the final number of clusters. 20NewsGroup is a good example. Secondly, K-means can deal with data with high dimensions, so we do not need to reduce the dimension of the data if we use k-means. However, at the same time, K-means also have some defects. For some complex data sets, the simple algorithm of K-means cannot achieve good results. For instance, when documents are distributed in space with complex shapes such as the Moon shape, k-means results will be unsatisfactory.

\subsection{Cluster description (TF-RDF)}

In this phase, we propose a novel technique TF-RDF to calculate the importance score of each word in a document, in this way we could get first n words as the topic of a cluster.
Before introducing TF-RDF, let us discuss the flaw of TF-IDF and C-TF-IDF method. TF-IDF is a useful method to evaluate the importance of a word in document. 

\begin{equation}
\text{TF-IDF}(t, d, D) = \text{TF}(t, d) \times \text{IDF}(t, D)
\label{equation1}
\end{equation}

where:
\begin{itemize}
    \item $\text{TF}(t, d)$ is the term frequency of term $t$ in document $d$.
    \item $\text{IDF}(t, D)$ is the inverse document frequency of term $t$ in the document set $D$, and is usually computed as:
    \begin{equation}
    \text{IDF}(t, D) = \log \left( \frac{|D|}{1 + |\{ d \in D: t \in d \}|} \right)
    \label{equation2}
    \end{equation}
    where $|D|$ is the total number of documents and $|\{ d \in D: t \in d \}|$ is the number of documents that contain the term $t$. The addition of 1 in the denominator is to prevent division by zero in the case where no document contains the term.
\end{itemize}

From its formulation(Equation~\ref{equation1} and Equation~\ref{equation2}), it is evident that the significance of a word within a document is determined both by its frequency of occurrence (as captured by TF) and the number of documents in which the term appears (as reflected by IDF).
The main purpose of IDF is to filter stop words, such as “the" and "of". These sort of terms almost appear in each of documents, and they should not be considered as a keywords of a document. Therefore, for these terms, TF-IDF will allocate them a low score. 

Nevertheless, when dealing with exceedingly large documents, the count of documents containing term t can surge abnormally. This phenomenon arises because, as the size of a document expands, the likelihood of any given word appearing across multiple documents increases. Under such circumstances, TF-IDF may falter in effectively filtering out stop words, simultaneously assigning unduly low scores to certain pertinent keywords. To illustrate, while a keyword like "Math" might be absent in some concise documents, it becomes implausible for a document exceeding 100,000 words to never reference the term "Math".

\begin{equation}
\text{C-TF-IDF}(t, c) = \text{TF}(t, c) \times \text{C-IDF}(t, A,f_t)
\label{equation3}
\end{equation}
\begin{equation}
\text{C-IDF}(t, A,f_t) = \log \left( 1 + \frac{A}{ f_t } \right)
\label{equation4}
\end{equation}

Where:
\begin{itemize}
    \item \( t \) represents a term.
    \item \( A \) represents average number of words per class
    \item \( f_t \) represents the frequency of term t across all classes
    \item \( TF(t,c) \) is the frequency of word t in class c.
    
\end{itemize}

Upon scrutinizing the C-TF-IDF formulations, as represented in Equations~\ref{equation3} and~\ref{equation4}, it becomes evident that for voluminous documents, the magnitude of the \( A \) parameter can burgeon significantly. Under such circumstances, the influence of \( f_t \) on the IDF score diminishes, relegating its role, and making the term frequency (TF) predominantly instrumental in determining the final C-TF-IDF score. Consequently, given the elevated term frequencies of certain stop words, they can be mistakenly attributed higher C-TF-IDF values. This skew leads to bias, potentially resulting in undesirable outcomes.

\begin{equation}
\text{TF-RDF}(t, d, \theta, n_t) = \text{TF}(t, d) \times \text{RDF}(t, d, \theta)
\label{equation5}
\end{equation}

where:
\begin{itemize}
    \item \( \text{TF}(t, d) \) represents the term frequency of term \( t \) in document \( d \), which is identical to the TF in TF-IDF.
    
    \item \( \text{RDF}(t, D) \) is the relative document frequency defined as:
    \begin{equation}
    \text{RDF}(t, d, \theta) = \log \left( \frac{\theta}{1 + n_{t,d}} \right)
    \label{equation6}
    \end{equation}
    where \( \theta \) is a hyperparameter and $n_{t,d}$ represents the count of occurrences of term $t$ in documents other than document $d$. The addition of 1 in the denominator is to avoid division by zero when no document contains the term.
\end{itemize}

In this work, we introduce TF-RDF(Term Frequency - Relative Document Frequency),as described by Equations~\ref{equation5} and \ref{equation6}, which modified the IDF component of TF-IDF so it could deal with large size document and properly filter the stop words. 

The primary concept behind TF-RDF hinges on employing \( n_{t,d} \) as a penalizing factor, with \( \theta \) serving as a modulating parameter to fine-tune the extent of this penalty. In essence, for a given term \( t \), its frequency across documents other than document \( d \) inversely impacts its RDF score. The underlying principle posits that if a term \( t \) is prevalently observed not just in document \( d \), it is less likely to signify a distinct topic. While this approach echoes the sentiment of IDF, RDF offers a relative measure rather than an absolute one. 

The TF-RDF method introduces a relative measure governed by the hyper parameter $\theta$. Essentially, $\theta$ sets a threshold for $n_{t,d}$: when term $t$ appears more than $\theta$ times outside document $d$, its TF-RDF score is effectively reduced (turning negative). Proper calibration of $\theta$ is crucial for optimal performance. A value too small for $\theta$ might inadvertently penalize terms that have a reasonable frequency across documents, skewing the TF-RDF scores in favor of rare words. Conversely, when $\theta$ is set excessively large, the model exhibits similar drawbacks as seen with C-TF-IDF. In this scenario, the influence of RDF becomes marginalized, causing term frequency (TF) to largely dictate the score. As a consequence, scores for stop words may be unduly inflated.

\subsection{The choice of $\theta$}

As previously discussed, the hyperparameter \(\theta\) serves as a threshold for the RDF value. Both excessively large and small values for \(\theta\) can adversely affect the performance of TF-RDF. This threshold is established to filter out stop words and enhance the ability of RDF to assign appropriate values to desired keywords. In determining this threshold, the distribution of word frequencies in a collection of documents must be considered. Stop words, given their nature, appear tens of thousands of times across document collections. Hence, it's essential to examine the frequency distribution of words in our target documents to set an appropriate threshold. A visual inspection using a histogram offers an effective method to assess this distribution.

\begin{figure}
\centering
\includegraphics[width=0.5\textwidth]{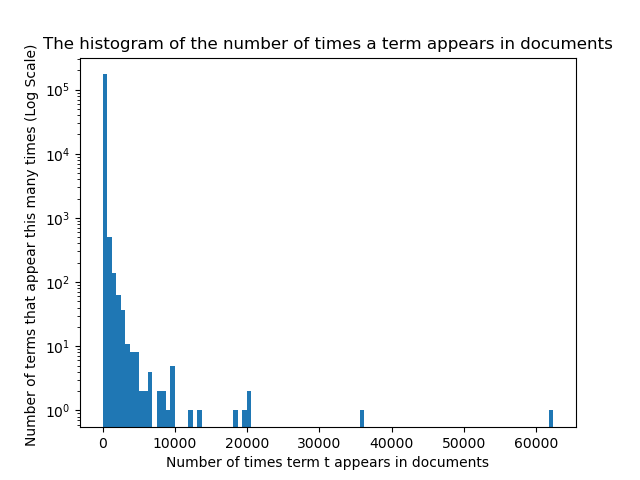}
\caption{\label{fig:figure1} We observe that the majority of terms appear fewer than 60 times, while certain stop words manifest tens of thousands of times. Given such a term distribution, our goal is to set the threshold $\theta$ to cover noise-free terms. Therefore, we recommend a $\theta$ value ranging from 5000 to 20000.}
\end{figure}
 
The selection of \(\theta\) offers a relative measure for TF-RDF. It's crucial to strike a balance where \(\theta\) is neither excessively large (risking the inclusion of noise) nor too diminutive (which might exclude significant keywords). A heuristic choice of this parameter, as exemplified by a value of 5000 in our case, can often yield satisfactory results. For those aiming for peak performance, conducting a grid search is advisable, factoring in the unique distribution properties of the given dataset.

\subsection{Visualization}

We must compress the final result into two dimensions in the visualization step. MPTopic still uses the UMAP~\cite{umap} algorithm for dimension reduction, and the parameters remain the same compared to our dimension reduction. The only difference is that we change the n\_component from 5 to 2 because we want the visualized data to be a 2-dimensional plane.

\subsection{Evaluation}
As highlighted in our methodology, our approach introduces two significant contributions. Firstly, we have devised an enhancement that elevates the quality of clustering outcomes. Secondly, we have proposed the TF-RDF, a novel technique designed to overcome the inherent limitations of both C-TF-IDF and TF-IDF. This advancement subsequently promotes the topic modelling performance of MPTopic.

For the purpose of rigorous evaluation, we employed a battery of metrics to appraise the clustering results. These metrics encompass: purity value, normalized mutual information (NMI), Rand index, adjusted mutual information (AMI), and adjusted Rand index~\cite{sklearn}. Furthermore, for evaluating the coherence of the topics generated, we utilized two topic coherence metrics.

\section{Experiments}
\subsection{Datasets}

The utilization of the purity value as an evaluative criterion necessitates a dataset with manual annotations, enabling a post-hoc assessment of label consistency within each cluster produced by the clustering algorithm. Given this prerequisite, we conducted our clustering experiments on an ensemble of seven distinct datasets: 20newsgroup, BBC news, AG news, Trec, Tweet(Emotion), Tweet(Emoji), and Yahoo Q\&A. For enhanced precision in our evaluations, each dataset was systematically partitioned into ten evenly sized subsets. Median values were computed from these subsets as a robust measure to mitigate potential outliers and anomalies. Subsequent analyses were carried out using these median values to discern the relative performance of various models across the datasets.

For the evaluation of topic descriptions, our experiments exclusively employed the 20newsgroup dataset. Given the intrinsic structure of the 20newsgroup dataset, which comprises exactly 20 classes, our directive was to configure all algorithms to yield 20 clusters. The topic coherence scores of these clusters served as the primary metric for evaluation.

\subsection{The evaluation of clustering algorithms}

\begin{table}
\centering
\includegraphics[width=1\textwidth]{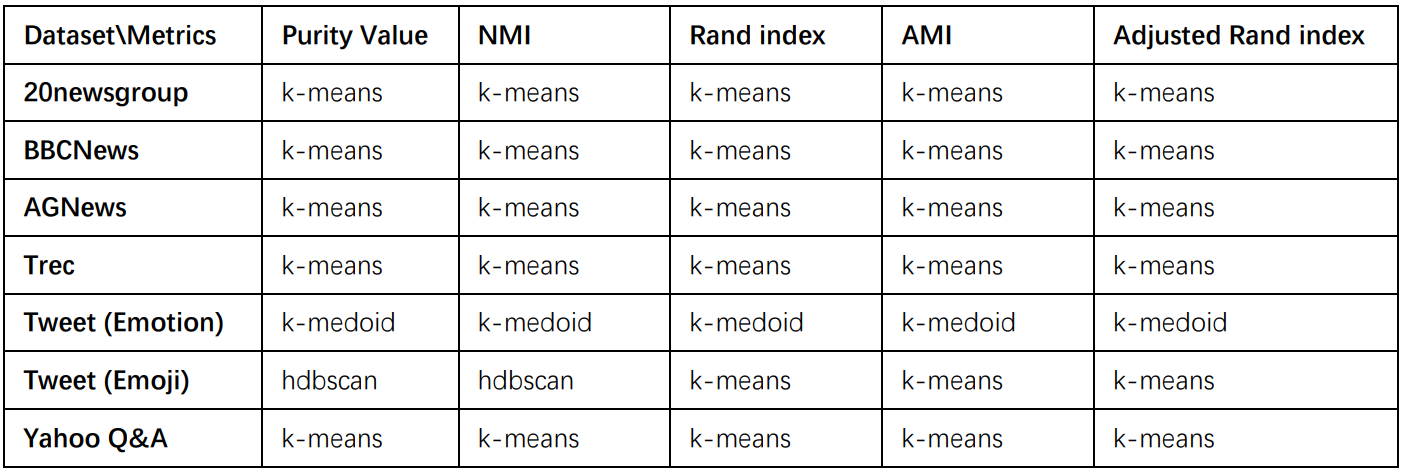}
\caption{\label{fig:mptopic_table}The Table shows the best clustering algorithm for different datasets and metrics. In most cases, the k-means algorithm has the best performance. }
\end{table}

In this section, we aim to demonstrate that HDBSCAN may not be the optimal choice for achieving superior clustering results when the number of classes is known or estimated. For the sake of this experiment, we maintained consistent configurations across all setups, with the sole exception being the clustering algorithm used. This approach enables us to isolate and observe the performance differences among various clustering methods.

The experiment was done on seven datasets(see Table~\ref{fig:mptopic_table}). The conclusion is that: On the Tweet(Emotion) dataset, the k-medoid algorithm gives the best performance. On the Yahoo Q\&A dataset, all algorithms have similar performance. On the rest of the datasets, k-means gives the best performance.

In this experiment, HDBSCAN is not performing well. This result supports why we want to provide the user with more options on the clustering algorithm: Sometimes, HDBSCAN is not the best solution.

\subsection{The evaluation of cluster description}

Topic modeling aims to discover suitable topics for a collection of documents, with each topic characterized by a set of keywords. In this study, we directly evaluate the final outputs of MPTopic, Bertopic, and Top2Vec. We employ the 20newsgroup dataset and instruct each of these algorithms to produce 20 topics. Subsequently, we assess whether these topics comprise clear, precise, and cohesive keywords.

\begin{table}[htbp]
\centering
\normalsize
\begin{tabular}{|l|c|c|}
      \hline
      \textbf{Method} & \textbf{TC(Pairwise)} &\textbf{TC(Centroid)}\\
      \hline
      MPTopic & \textbf{0.1105} & \textbf{0.6485} \\
      Bertopic & 0.0201 & 0.5711 \\
      Top2Vec &  0.0876 & 0.5834 \\
      \hline
\end{tabular}
\caption{
\footnotesize
Our MPTopic utilizes k-means for clustering and employs TF-RDF to characterize each cluster. This method exhibits superior performance, especially when there is precise knowledge or estimations of the number of clusters. Notably, Bertopic demonstrates an exceedingly low score on TC(Pairwise). This can be attributed to the limitations of its C-TF-IDF in handling large-sized documents. Specifically, when the number of clusters is small (and consequently the cluster size is large), many stop words are inaccurately assigned high scores.}
\label{tab:table2}
\end{table}

The Topic coherence (TC) is a metric to evaluate the quality of topics produced by a topic modelling. In essence, Topic Coherence examines the coherence or similarity between words within a topic.

In more detail, when a topic model presents us with a topic constituted of a set of words, we expect these words to be highly related and representative of a clear concept or subject. For instance, for a topic about sports, we might expect words like "ball," "game," and "team" rather than unrelated terms like "space," "planet," or "rocket."

In our approach, we utilize Word2Vec~\cite{word2vec} to convert each keyword within topics into a vector representation. Subsequently, we compute the cosine similarity between pairs of terms and also between each term and their centroid. By evaluating these two types of TC scores, we obtain a metric that reflects the quality of topics generated by each algorithm. Given that the score is derived from cosine similarity, its value range lies between -1 and 1.  Utilizing Word2Vec embeddings for evaluation presents a robust approach. Even though both MPTopic and the baseline algorithms employ deep learning methods for word embeddings, neither adopts Word2Vec specifically. Thus, leveraging Word2Vec as an impartial referee for evaluating topic modeling performance can be deemed a balanced and judicious solution.

\section{Conclusion}
\subsection{Summary}

In this paper, we introduce MPTopic, an innovative topic modeling algorithm designed to overcome the limitations of Top2CVec and BERTopic, particularly when users seek to specify the exact number of topics. At the core of MPTopic is TD-RDF, a robust text representation technique. By integrating TD-RDF with a range of clustering algorithms—including k-means, k-medoids, HDBSCAN, and Birch—we aim to offer greater versatility in topic extraction. Notably, the combination of MPTopic with k-means demonstrates superior performance compared to baseline models in both clustering outcomes and topic modeling results.

\bibliographystyle{splncs04}
\bibliography{paper}

%




\end{document}